\documentstyle[prl,aps,amsfonts,amssymb,epsfig,floats]{revtex}
\begin{document}
\draft
\preprint{}
%
\twocolumn[\hsize\textwidth\columnwidth\hsize\csname
@twocolumnfalse\endcsname
\title{$n(\vec{k})$ Structure and Matrix Elements in Photoemission:
Results from Insulating Ca$_2$CuO$_2$Cl$_2$}

\author{F. Ronning, C. Kim, K.M. Shen, N.P. Armitage, A. Damascelli,
D.H. Lu, and Z.-X. Shen}
\address{
Department of Physics, Applied Physics and Stanford Synchrotron
Radiation Laboratory,\\ Stanford University, Stanford, CA 94305,
USA}
\author{L.L. Miller}
\address{Department of Physics, Iowa State University, Ames Iowa, 50011}

\date{\today}
\maketitle
\begin{abstract}
The photon energy dependence of angle-resolved photoemission data on
Ca$_2$CuO$_2$Cl$_2$ is presented, and the role of the matrix element
is investigated.  It is shown that despite strong variations in the
matrix element, the loss in intensity as one crosses the
antiferromagnetic zone boundary is a robust feature of the Mott
insulator.  As expected for a two dimensional system, the dispersion
remains independent of the photon energy even when the matrix element
does not.

\end{abstract}
%
\vskip2pc]
\narrowtext

The question of how a metal evolves into an insulator is one of the
most fundamental in solid state physics.  For non-interacting
electrons, the Fermi surface shrinks and eventually disappears as a band is
filled.  When electron-electron interactions dominate, the situation
is less clear.  By considering the strong Coulomb interaction, Mott
qualitatively described how a material predicted by band theory to be
a metal would in fact be an insulator.\cite{Mott} However, it remains
unclear as to how the details of the electronic structure evolves from
a half-filled metal to a Mott insulator.  One of the most
characteristic features of any metal is its Fermi surface which can be
defined by the surface of steepest descent in the electron momentum
distribution function, $n(\vec{k})$.  Thus an equivalent question to
the one above is how does the Fermi surface of a metal vanish as
strong electron correlations drive the system into an insulator?

In the context of specific many-body models such as the Hubbard model,
it has been shown that a structure in $n(\vec{k})$ survives even when
the on site Coulomb U drives the system insulating, albeit the
discontinuity in $n(\vec{k})$ which existed in the metal has been
washed out.\cite{Dagotto} This effect is linked to the fact that
$n(\vec{k})$ reflects the underlying Fermi statistics of the
electronic system.  For the specific case of a two dimensional square
lattice that resembles the CuO$_{2}$ planes of the cuprates, there is
a drop in $n(\vec{k})$ across a line that is close to the
antiferromagnetic zone boundary.\cite{Dagotto}

It is believed that the information under the sudden approximation of
the momentum distribution function, $n(\vec{k})$, can be extracted
from angle-resolved photoemission spectroscopy (ARPES) data via the
relation $n(\vec{k})$ = $\int A(\vec{k},\omega)f(\omega) d\omega$.
\cite{Randeria} Here $A(\vec{k},\omega)$ is the spectral function, and
$f(\omega)$ is the Fermi function.  In a real experiment, the
$n(\vec{k})$ structure would be further modified by the
photoionization cross-section.  In the metallic state of optimally
doped Bi$_2$Sr$_2$CaCu$_2$O$_{8+\delta}$(BSCCO) the steepest descent
of $n(\vec{k})$ gives a Fermi surface consistent with traditional
ARPES analysis methods, despite the complication of matrix elements.
The intriguing new result is that the $n(\vec{k})$ pattern of the
insulator, Ca$_2$CuO$_2$Cl$_2$(CCOC), is strikingly similar to the
$n(\vec{k})$ pattern seen in BSCCO.\cite{Science} This realization,
coupled with many-body theoretical calculations on various forms of
the Hubbard model\cite{Dagotto}, suggests that the insulator pattern
contains information that is related to a Fermi surface which has been
destroyed by strong electron-electron interactions thus giving a
qualitative concept of a remnant Fermi surface as the surface of
steepest descent in $n(\vec{k})$.  Although the detailed shape of the
remnant Fermi surface is not crucial, it acts to emphasize a robust
feature which we observe in the insulator.  While this may not be a
rigorous definition, as the Fermi surface is only defined for a metal,
this idea allows a practical connection from the pseudogap seen in
underdoped cuprates to the properties of the insulator.\cite{Laughlin}

A photon energy dependence study of this issue is important to extract
the intrinsic $n(\vec{k})$ structure from the raw data which can be
affected by the matrix element.  Because the single particle spectral
function obtained from the half filled CCOC insulator will not suffer
from possible complications arising due to intrinsic electronic
inhomogeneity, bilayer splitting, or superstructure as in other
cuprates, this study may be able to provide insight on some of the
recent controversy regarding the Fermi surface seen in BSCCO.

In this report we show that despite the aberrations caused by the
matrix element effect, the drop in $n(\vec{k})$ as one crosses the
antiferromagnetic zone boundary is a robust feature of CCOC. This
result differs from the conclusion of a recent study based on data
from only a single cut across the magnetic zone
boundary.\cite{Haffner} We also show that the dispersion remains
independent of photon energy even when the matrix element does not.
This is an expected result from an ideal two dimensional system.

The measurements are conducted at Beamline V-3 of the Stanford
Synchrotron Radiation Laboratory (SSRL).  The angular resolution is
$\pm 1^{\circ}$.  The CCOC samples\cite{Lance} are oriented prior to
the experiment by the Laue method and are cleaved {\sl in
situ}.\cite{cleavecomment} The light is incident at $45^{\circ}$ to
the surface normal and has an in-plane component of the polarization
along $k_{x}$, which is also parallel to the Cu-O bonds.  The chamber
pressure is better than $5\times10^{-11}$ torr.  Data is taken in the
$\vec{k}$-space octant $(0,0)-(\pi,0)-(\pi,\pi)-(0,0)$ and at
temperatures ranging from 100 to 185K, well below the N\'eel
temperature of 250K. The insulating samples are checked for charging
by varying the photon flux.  All data are normalized by the photon
flux which is measured by a Au mesh upstream from the sample.
\begin{figure}[t!]
\centerline{\epsfig{figure=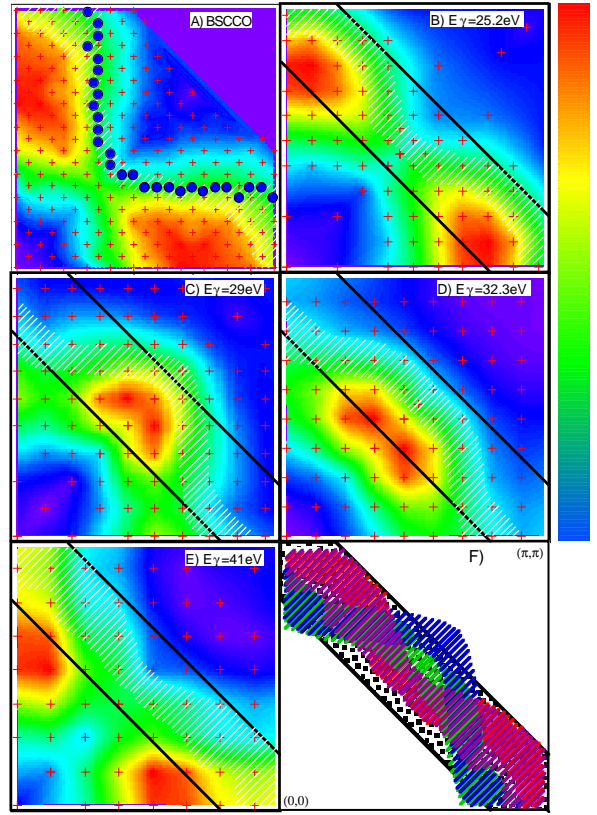,width=7.8cm,clip=}}
\vspace{.2cm} \caption{(color) Integrated spectral weight.  The
crosses indicate where spectra were taken.  The data is
symmetrized about the $k_{x}$=$k_{y}$ line.  A color scale is
given on the right. (A) optimally doped BSCCO. The white hashed
region indicates the approximate location of the Fermi surface
determined from $n(\vec{k})$.  The dots illustrate the position of
the Fermi surface as determined by the traditional method for
analyzing ARPES data.  (B) CCOC shows a striking similarity of the
insulator to the metal allows the identification of the white
hashed region as a remnant Fermi surface. Comparison of (B)
through (E) show CCOC taken at photon energies of 25.2, 29, 32.3,
and 41eV. The intensity maxima varies between different panels,
but the loss of intensity as one approximately crosses the
antiferromagnetic zone boundary is a consistent feature.  (F)
shows the remnant Fermi surface determined from (B) through (E).
The shaded region gives an estimate for the uncertainty in the
remnant Fermi surface.  The boundary of this shaded region is
drawn with black lines on panels (B) through (E).} \label{Figure1}
\end{figure}

We begin with a short discussion of matrix element effects in
photoemission.  Extracting the single particle spectral function from
ARPES measurements is complicated by the fact that the measured
photoemission intensity is a product of the spectral function and the
matrix element.  In interacting electron systems, it is impossible to
calculate the matrix element exactly, thus further complicating the
ARPES analysis.  In general it is a function of the experimental
geometry, photon energy, and the electronic wave function.  However,
symmetry arguments can be very powerful in understanding some of the
general properties of the matrix element.  Since its details are not
well understood, the objective in a given photoemission study must be
to focus on only those features of the data which are robust against
variations in the matrix element.

For the $n(\vec{k})$ plots on CCOC, the spectra are integrated from
-0.5 to 0.3 eV binding energy relative to the valence band maximum.
Spectra are taken at the crosses, and the data is symmetrized about
$k_{x}$=$k_{y}$ to better illustrate the remnant Fermi surface.
Figure 1 shows the original comparison between metallic BSCCO and
insulating CCOC in panels (A) and (B).\cite{Science} In both cases the
geometry of the experimental setup is identical.  Assuming a
$d_{x^{2}-y^{2}}$ orbital symmetry we would expect a vanishing cross
section as one approaches the line $k_{x}$=0.  This is observed in
both data sets.  In BSCCO the only drop in intensity which is not
naturally explained by matrix elements is where a Fermi surface
crossing has occurred.  As seen in the figure the intensity drop
matches the traditional method for determining a Fermi surface
crossing by following the dispersion by eye is indicated by the dots.
A similar drop in intensity is observed in the insulator.  Although
the feature is less well defined here, the striking resemblance it
bears to the metal suggests that the origin is similar, and hence it
is qualitatively described as a remnant Fermi surface.  This allows us
to make a natural link between the dispersion in the insulator and the
pseudogap in the underdoped regime of the high-T$_{c}$
cuprates.\cite{Science}\cite{Laughlin}

As mentioned above, we must still show the robustness of this feature
to demonstrate the validity of this concept.  Panels B through E of
Figure 1 show $n(\vec{k})$ patterns of CCOC taken at photon energies
of 25.2, 29, 32.3, and 41eV. At first glance they may appear
different.  One clearly sees the position of maximum intensity shift.
Thus the position of this maximum along the antiferromagnetic zone
boundary can be attributed to matrix element effects.  However, the
expected suppression for a state with $d_{x^{2}-y^{2}}$ symmetry as
one approaches the zone center is observed in all the data.  In fact,
a careful look at the data shows that they are very similar.  The
exact shape of the remnant Fermi surface may change slightly, but at
all photon energies used there is a loss of spectral weight as one
crosses the approximate AF zone boundary from $(0,\pi)$ to $(\pi,0)$.
Panel F plots the remnant Fermi surface as determined by figures 1B
through 1E. The shaded region in F gives a rough estimate for the
shape and uncertainty of the remnant Fermi surface.  Thin black lines
in panels B through E form the border of the shaded region shown in F.
It may appear that the remnant Fermi surface is more hole like or
electron like depending on the photon energy chosen, but the broadness
and the minor variability due to matrix elements prevent one from
clearly identifying the remnant Fermi surface as either case, as
illustrated by the hashed region.  However, the point which is clear
from the data is that globally, there is always an unexpected loss in
intensity as one crosses the region in the Brillouin zone indicated by
the shaded area in figure 1F, which represents the remnant Fermi
surface.  Indeed, the remnant Fermi surface is a robust feature of the
insulator.
\begin{figure}[t!]
\centerline{\epsfig{figure=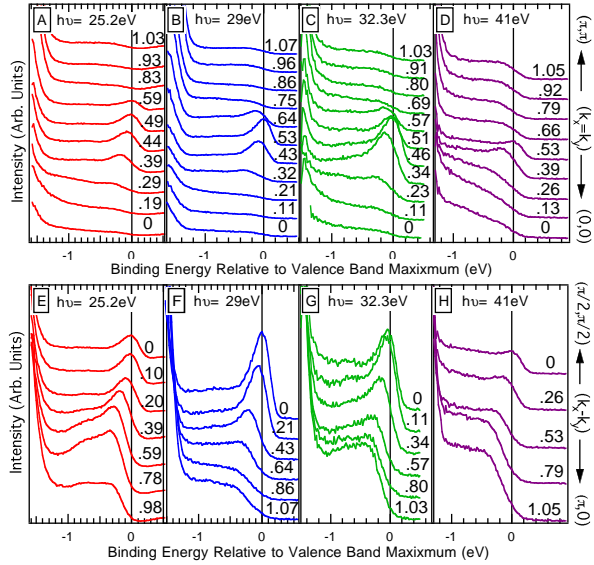,width=7.8cm,clip=}}
\vspace{.2cm} \caption{(color) ARPES spectra along 2 cuts, (0,0)
to $(\pi,\pi)$ and $(\pi,0)$ to $(\pi/2,\pi/2)$, and 4 different
photon energies (25.2, 29, 32.3, and 41eV).  The intensities of
the features vary, but the dispersion remains the same.}
\label{Figure2}
\end{figure}

Although the $n(\vec{k})$ image plots can provide a wealth of
information and are extremely good for summarizing data, it is
important to look at the raw data to fully appreciate what information
is being given by the image plots.  This is done in Figure 2.  Panels
a through d (e through f) plot the EDC's from (0,0) to $(\pi,\pi)$
($(\pi,0)$ to $(\pi/2,\pi/2)$) taken at 25.2, 29, 32.3, and 41eV
photon energy respectively.  We find the spectra are qualitatively
similar.  This is true even at 41eV where the peak is poorly defined
throughout the zone.  To examine them more closely, Figure 3 plots
both the dispersion of the peak position and $n(\vec{k})$ together for
the (0,0) to $(\pi,\pi)$ cut.  Only slight differences exist between
the 4 photon energies.  It is apparent that the matrix element can be
responsible for shifting the maximum in intensity relative to the
maximum in dispersion.  Such an effect is seen between the 29 and 41eV
data, which may be rigorously compared.  A similar shift of spectral
weight has been reported by Haffner {\sl et al.} in
Sr$_2$CuO$_2$Cl$_2$.\cite{Haffner} These authors conclude that there
is no $n(\vec{k})$ structure or remnant Fermi surface.  We believe the
intensity shift is a small modulation due to the a matrix element
compared with the robust underlying $n(\vec{k})$ structure seen
throughout the zone at all photon energies.  To illustrate this point
we also plot the spectra, dispersion, and $n(\vec{k})$ along $(\pi,0)$
to $(0,\pi)$.  From the raw spectra in figure 2 and the peak positions
shown in figure 3 it is clear that the dispersion does not change.
Meanwhile the intensity of the feature varies randomly along this cut.
In the extreme case between 25.2eV and 29eV the intensity is
increasing as one approaches $(\pi/2,\pi/2)$ for the former, and
decreasing for the latter.  This completely contrasts the situation
along (0,0) to $(\pi,\pi)$, where there are only slight shifts of the
intensity profile, but the global structure remains the same.
\begin{figure}[t]
\centerline{\epsfig{figure=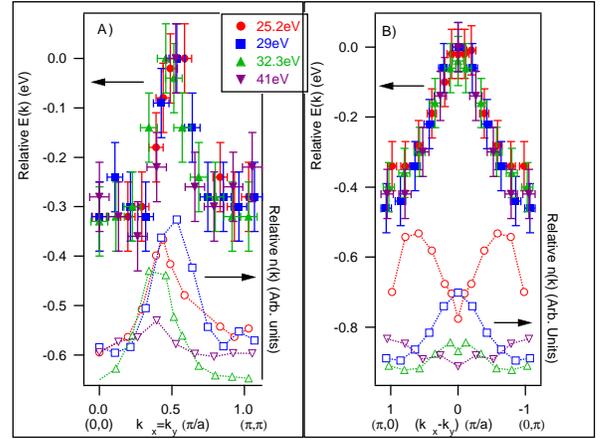,width=7.8cm,clip=}}
\vspace{.2cm} \caption{(color) Plots the peak position and
$n(\vec{k})$ from the spectra in figure 2.  For both cuts the
dispersion is independent of photon energy.  The data along
$(\pi,0)$ to $(0,\pi)$ has been symmetrized.  Along (0,0) to
$(\pi,\pi)$ the intensity has minor shifts, but is overall
consistent indicating robustness of the remnant Fermi surface.
For $(\pi,0)$ to $(\pi/2,\pi/2)$ the intensity varies randomly
indicating that this modulation in intensity is due to matrix
elements.  Filled symbols indicate $E(\vec{k})$ while open symbols
are for $n(\vec{k})$.} \label{Figure3}
\end{figure}

The cuprates are generally believed to be two dimensional electronic
systems.  In such an ideal case, the electronic states probed by ARPES
are independent of photon energy, and hence $A(\vec{k},\omega)$
extracted from the spectra will remain the same.  Thus it is natural
to question this assumption when photon energy dependent shifts in the
physics are reported.  This may seem particularly true for the half
filled cuprates which possess three dimensional long range magnetic
order.  However, in figures 2 and 3 we have shown that the dispersion
of the insulator is essentially independent of photon energy.  Three
dimensionality may still play a role in causing a small shift in
dispersion with photon energy, but it is
a safe assumption to treat the electronic structure of the half filled
insulator as basically two dimensional.

Recently, the role of the matrix element effects is also under
examination regarding the Fermi surface of
BSCCO.\cite{elecBSCCOFS}\cite{BSCCOphotdep}\cite{Bansil} Several
groups have reported the existence of an electron-like Fermi surface
centered at (0,0).\cite{elecBSCCOFS} These groups report a subtle
change of dispersion near the flat band region around $(\pi,0)$,
causing a change in the shape of the Fermi surface.  Largely due to
the difficulty in understanding the effects caused by photoelectron
matrix elements, there is no concensus in attempting to resolve this
issue.\cite{BSCCOphotdep} The features in BSCCO are better defined
than the relatively broad features in CCOC, and one might suspect
that, as a result, they may be more susceptible to photon energy
dependence.  However, the idea that the dispersion should be
independent of photon energy for an ideal two dimensional system is the
same, regardless of the resolution.  A change in the dispersion can
not be a trivial matrix element effect.  Possible explanations for a
dispersion which changes with photon energy include: dispersion in
$k_{z}$, bi-layer splitting, and phase separation.  With the exception
of Pb doped samples BSCCO also has the additional complication of
superstructure in the Bi-O planes.  In BSCCO, it would be surprising
for the dispersion in $k_{z}$ to have a large effect considering the
two
dimensional nature of the cuprates.  In reality though it is likely
that all of these effects play roles which result in a potentially
very complex picture.

Interestingly, models which have multiple component electronic
structures where the wavefunctions and dispersions differ could have a
simple explanation for the observed controversy.  In such a case,
variations in the matrix element could now lead to different
interpretations of the observed ARPES spectra at different photon
energies.  This is possible because there may now be a different
matrix element for each component, and each matrix element may vary
differently with photon energy.  Thus causing one component to be more
dominant at one photon energy while hidden at another.  This
would naturally lead to different interpretations as a function of
photon energy.

We have shown that the loss in intensity as one crosses the
antiferromagnetic zone boundary is a robust feature of the insulator,
Ca$_2$CuO$_2$Cl$_2$, which can not be explained solely by matrix
element effects.  However, the photon energy dependence does
underscore the qualitative rather than quantitative nature of the
remnant Fermi surface concept.  We argue that much physics can be
learned in spite of the effect which matrix elements can have in
ARPES, as long as care is taken to properly sort out the intrinsic
versus the extrinsic physics.  In particular, the resulting connection
between the d-wave like dispersion and the pseudogap in the underdoped
regime is robust.

We thank C. D\"urr, M.S. Golden, and J. Fink for very open and
stimulating discussions. This work was performed at Stanford
Synchrotron Radiation Laboratory which is supported by the DOE Office
of Basic Energy Science, Division of Chemical Sciences and Material
Sciences. It is also supported by the ONR grant N00014-98-1-0195 and
NSF grant DMR-9705210.


\begin{references}
\bibitem{cleavecomment} Data taken with 29
and 41eV photon energy are from the same cleave, while data
presented at other photon energies were taken on separate cleaves.
Due to the fact that there will always be a small uncertainty in
aligning a sample, only data from the same cleave may be compared
when looking for small shifts.  However, the effects we are
interested in are robust features which are independent of small
shifts in energy or k-space.  So all the data presented here is
valid for comparison in the scope of this report.
\end{references}
\end{document}